\documentstyle[sprocl,epsf]{article}
\scrollmode
\begin{document}
\title{STRING SOLITONS AND BLACK HOLE THERMODYNAMICS}
\author{Ramzi R. Khuri}
\address{Department of Physics, McGill University\\ Montreal,
PQ H2X 2L6 Canada}
\maketitle \abstracts{ 
We discuss the role of string solitons in duality and 
examine the feature of compositeness, which allows for 
the interpretation of general solutions as bound states 
of supersymmetric fundamental constituents. This feature 
lies at the heart of the recent success of string theory 
in reproducing the Bekenstein-Hawking black hole entropy 
formula. Talk given at 19th annual MRST meeting, Syracuse, 
NY, May 12-13, 1997. McGill 97-20. }

The standard model of elementary particle physics has proven 
very succesful 
in describing three of the four fundamental forces of nature. 
In the most
optimistic scenario, the standard model can be generalized to 
take the
form of a grand unified theory (GUT), in which quantum 
chromodynamics 
(QCD), describing the strong force, and the electroweak theory, 
unifying
the weak interaction with electromagnetism, are synthesized 
into a single
theory in which all three forces have a common origin. 

The framework for
studying these three forces is that of Yang-Mills gauge theory,
a certain class of quantum field theory based on the principle 
of gauge
symmetry. In any quantum-mechanical theory, the natural length 
scale
associated with a particle of mass $m$ (such as an elementary 
particle)
is given by the Compton wavelength $\lambda_C=\hbar/mc$, where 
$\hbar$
is Planck's constant divided by $2\pi$ and $c$ is the speed of 
light.
Scales less than $\lambda_C$ are therefore
unobservable within the 
context of 
the quantum mechanics of this elementary particle.

Quantum field theory, however, has so far proven unsuccessful in 
describing
the fourth fundamental force, namely gravitation. The succesful 
framework  
in this case is that of general relativity, which, however, 
does 
not
seem to lend itself to a straightforward attempt at 
quantization. 
The 
main problem in such an endeavour is that the divergences 
associated with
trying to quantize gravity cannot be circumvented (or 
``renormalized")
as they are for the strong, weak and electromagnetic forces. 

Among the most interesting objects predicted by general 
relativity are 
black holes, which represent the endpoint of gravitational 
collapse. 
According to relativity, an object of mass $m$ under the 
influence of only 
the 
gravitational force ({\em i.e.} neutral with respect to the 
other three 
forces) will collapse into a region of spacetime bounded by a 
surface,
the event horizon, beyond which signals cannot be transmitted 
to an outside 
observer. The event horizon for the simplest case of a 
static, spherically
symmetric black hole is located at a radius $R_S=2 G m/c^2$, 
the 
Schwarzschild radius, from the collapsed matter at the 
center of the sphere,
where $G$ is Newton's constant. 
      
\def\buildchar#1#2#3{{\null\!
   \mathop{\vphantom{#1}\smash#1}\limits%
   ^{#2}_{#3}%
   \!\null}}
\def\ha{{1\over 2}}
  
In trying to reconcile general relativity and particle 
physics, even at an
intuitive level, a natural question to ask is whether 
they have a common
domain. This would arise when an elementary particle 
exhibits features 
associated with gravitation, such as an event horizon. 
This may occur 
provided 
$\lambda_C \ { } 
\lower3pt\hbox{$\buildchar{\sim}{<}{}$} \ { } R_S$,
which implies that, even 
within the framework of quantum mechanics, 
an event horizon for an 
elementary
particle may be observable. Such a condition is equivalent to
$m \ { } \lower3pt\hbox{$\buildchar{\sim}{>}{}$} \ { } m_P=
\sqrt{\hbar c/G} \sim 10^{19} GeV$, 
the Planck mass, or 
$\lambda_C \ { } 
\lower3pt\hbox{$\buildchar{\sim}{<}{}$} \ { } l_P
=\sqrt{\hbar G/c^3}$,
the Planck scale. 
It is in this domain that one may study a theory
that combines quantum mechanics and gravity, the so-called
{\em quantum gravity} (henceforth we use units in which
$\hbar=c=1$).

A problem, however, arises
in this comparison, because most black holes are thermal
objects, and hence cannot reasonably be identified with
pure quantum states such as elementary particles. In fact,
in accordance with the laws
 of {\em black hole thermodynamics} \cite{Bek,Hawk}\ 
black holes radiate with a (Hawking) temperature constant
over the event horizon and proportional to the surface gravity:
$T_H\sim \kappa$. Furthermore, black holes possess an entropy
$S =A/4G$, where $A$ is the area of the horizon (the area
law), and 
$\delta A \geq 0$ in black hole processes. Finally, the energy
law of black hole thermodynamics takes the form
$dM=(\kappa/8\pi) dA + \Omega_H dJ$, in analogy with 
$dE=TdS-PdV$, where in the former case $\Omega_H$ 
is the angular velocity of the horizon and $J$ the angular 
momentum of the black hole. 

In this framework, then, a pure state with $S=0$ such as an 
(electrically charged) elementary particle corresponds 
to a black hole with zero area. Such black holes arise 
as extremal limits of two-parameter (mass and charge)
families of charged black holes. From the cosmic censorship
principle, which forbids the existence of ``naked''
singularities ({\it i.e.} singularities not hidden behind a 
horizon), such solutions are required to satisfy a bound 
between the mass $M$ and charge $Q$, {\it e.g.}, $M \geq Q$.
Alternatively, this bound can be expressed in terms of
the outer (event) horizon $R_+$ and inner horizon $R_-$ 
of the black hole
(the latter horizon representing the limit of energy extraction 
from the black hole): $R_+ \geq R_-$. Extremality then
represents the saturation of the bound by $M=Q$ or
$R_+=R_-$. We shall see later that, while some extremal
black holes possess zero area, and therefore zero entropy,
others do not. The former can potentially correspond to 
elementary particles (pure states), while the latter 
correspond to an {\it ensemble} of particles or states.   

At the present time, string theory, the theory of one-dimensional
extended objects, is the only known reasonable
candidate theory  of quantum gravity. The divergences inherent 
in trying to quantize point-like gravity seem not to arise in
string theory. Furthermore, string theory has the potential
to unify all four fundamental forces within a common framework.

The two-dimensional worldsheet swept out by a string is embedded
in a higher-dimensional (target) spacetime, which in turn 
represents a 
background
for string propagation. At an intuitive level, one can see 
how point-like
divergences may possibly be avoided in string theory by 
considering
the four-point amplitudes arising in string theory \cite{GSW}). 
Unlike those of field theory,
the four-point amplitudes in string theory do not have 
well-defined
vertices at which the interaction can be said to take place,
hence no corresponding divergences associated with the zero size
of a particle. A simpler way of saying this is that the finite 
size of 
the string smooths out the divergence of the point particle.

The ground states of string theory correspond to conformal 
invariance
of the two-dimensional sigma model of a genus zero (sphere) 
worldsheet.
Solving the beta-function equation of this sigma-model then 
corresponds
to classical solutions of string theory. Within this classical 
theory,
the perturbative parameter is $\alpha'=1/(2\pi T_2)=l_s^2$, 
where $T_2$ is
the tension of the string and $l_s$ is the string length.

Perturbative quantum corrections in string theory take the 
form of 
an expansion in the genus of the worldsheet, with coupling 
parameter 
$g=\exp{\phi}$, where $\phi$, the dilaton, is a dynamical scalar
field.

Consistent, physically acceptable string theories possess 
supersymmetry
between bosons and fermions, and supersymmetric string theory 
requires
a ten-dimensional target space. This leads to another feature 
of string
theory, namely, compactification, {\em i.e.}, the splitting 
of the
ten dimensional vacuum into the product of
a four-dimensional vacuum and
a compact six-dimensional manifold which may be shrunk to a 
point. 

Finally, the property of string theory that is of most interest 
to us in these lectures is that of {\em duality}. At the 
simplest
level, duality is a map that takes one theory into another 
theory (or possibly
the same theory in a different domain). An immediate consequence 
of duality
is that the two theories are physically equivalent in the 
appropriate
domains. It then follows that calculations performed on one side
can be immediately carried over into the other, even if direct
calculations in the latter theory may have previously been 
intractable.

The two most basic dualities in string theory are the target 
space $T$-duality \cite{Tduality}\ and the 
strong/weak coupling $S$-duality \cite{Sduality}. 
Suppose
in a compactification one of the dimensions of the 
six-dimensional
compactification manifold is wrapped around a circle with radius 
$R$. Then $T$-duality is a generalization of a map that
takes a string theory with radius
$R$ into a theory with radius $\alpha'/R=l_s^2/R$. This 
implies that
a radius smaller than the string scale is equivalent to a radius
larger than the string scale. Effectively, then, the string 
scale
is a minimal scale, which conforms to our previous intuition 
that
the size of the string smooths out the point-like divergence. 
$T$-duality is a classical, 
worldsheet duality and in various compactifications has been 
shown
to be an exact duality in string theory. 

$S$-duality, by contrast, is a quantum (string loop), spacetime 
duality and  generalizes the map that takes the string coupling 
constant $g$ to its inverse $1/g$. Such a map takes the weak 
coupling 
domain into the strong coupling domain within a given string 
theory 
and allows us to use perturbative results in the latter. Also 
unlike
$T$-duality, $S$-duality has only been established exactly in 
the 
low-energy limit of string theory. 

These
two dualities and the interplay between them
are at the heart of the recent activity in string theory.
This activity has also been fueled by the realization that 
perturbative string theory is insufficient
to answer the most fundamental questions of string theory, such 
as vacuum selection, supersymmetry breaking, the cosmological
constant problem and, finally, the problem of understanding
quantum gravity from string theory. All these questions require
{\em nonperturbative} information. 

What kind of objects arise in nonperturbative physics? 
Solitons, or topological defects \cite{Raj}, 
are inherently nonperturbative solutions, representing
objects with mass $m_s \sim 1/g^2$, where $g$ is the coupling
constant of the theory. Examples of solitons are magnetic 
monopoles or domain walls. The connection between duality and 
solitons often involves the interchange of perturbative,
fundamental (electric) particles with nonperturbative, solitonic
(magnetic) objects. This is the main feature of the 
Montonen-Olive
conjecture \cite{Mont}\ for $N=4$, $D=4$ supersymmetric
Yang-Mills gauge theory, which postulates the existence of a 
dual version of the theory in which electric gauge bosons
and magnetic monopoles interchange roles. In this scenario,
the monopoles become the elementary particles and the
gauge bosons become the solitons. 

What does a duality map look like? Let us look at the simplest
example. Consider four-dimensional point-like electromagnetism,
a $U(1)$ gauge theory with gauge field $A_M$ and field strength
$F_{MN}=\partial_M A_N - \partial_N A_M$.  
The field of an electric charge $Q_e$ located
at the origin is given by $E_r=F_{0r}=Q_e/r$,
where $r$ is the radial coordinate.
The field of a magnetic charge $Q_m$ located
at the origin is given by $B_r=F_{\theta\phi}=Q_m/r$,
where $\theta$ annd $\phi$ are coordinates on the
two-sphere $S^2$.
Now the {\it dual} of the field strength is given by
$\tilde F_{AB}=(1/2) \epsilon_{ABCD} F^{CD}$, where 
$\epsilon_{ABCD}$
is the four-dimensional Levi-Civita tensor. This map 
is easily generalized to an arbitrary number of dimensions.
In all cases, the dual of the dual of a tensor reproduces the
original tensor up to a sign: $\tilde{\tilde F} = \pm F$.
For our four-dimensional example,
$\tilde F_{0r}=F_{\theta\phi}$. So under the map that takes
$F$ to $\tilde F$, there is an interchange of $Q_e$ and $Q_m$.
Another way of saying this is that, in the dualized version of
electromagnetism, what had previously appeared as electric charge
now appears as magnetic and vice-versa.

Now consider a one-dimensional extended object, a string.
In analogy with the point-particle, the string couples to
an antisymmetric gauge field, but this time in the form of
a two-tensor $B_{MN}$, with three-form field strength $H_{MNP}$.
For the dual of $H$ to represent the field strength of a string,
the Levi-civita tensor must be six-dimensional:
$\tilde H_{ABC}=(1/6) \epsilon_{ABCDEF} H^{DEF}$. So a
duality between string theories is most naturally formulated
in six dimensions, where the ``electric'' string charge   
read from $H_{01r}$ (where $x^1$ is the direction of the string)
is interchanged with the ``magnetic'' string charge read
from $H_{\chi\theta\phi}$, 
where $\chi$, $\theta$ and $\phi$ are coordinates on the
three-sphere $S^3$.

In the low-energy limit, the main feature
of string/string duality is the following:
in one version of string theory, there exists 
an electric, elementary string solution corresponding
to a perturbative state of the theory and a magnetic,
solitonic string solution corresponding to a nonperturbative
state. In the dual string theory, the solution that appears
electric in the first version now appears magnetic and vice-versa,
while the state that appears perturbative in the first version
now appears nonperturbative and vice-versa. The duality map
relates the string coupling $g$ of the first version to
that of its dual $g'$ via $g=1/g'$, so that the weak and 
strong coupling regimes of the two theories are interchanged. 
A interesting and nontrivial consequence of string/string
duality is that, in compactifying down to four dimensions,
the duality map takes the spacetime,
strong/weak coupling $S$-duality of one version into
the 
worldsheet, target space $T$-duality of the dual version. 
Since $T$-duality is in many cases established as exact,
the conjectured $S$-duality
would then follow as a consequence of string/string duality.

{}For the purpose of understanding black hole thermodynamics
from string theory, however, the most important feature of 
duality is that, by applying the various
duality maps, one can construct spectra of electric,
magnetic and dyonic states, also represented by classical
solutions of string theory \cite{Prep}. 
Using the solutions/states correspondence,
we compare the Bekenstein-Hawking entropy obtained from the
area of the classical solution to the quantum-mechanical
microcanonical counting of ensembles of states. 

It turns out that solutions in string theory possess a very nice
feature that greatly facilitates this comparison, namely,
that of {\it compositeness}, whereby arbitrary
solutions arise as bound states of single-charged
fundamental constituents. For example, 
the Reissner-Nordstr\"om solution
of Einstein-Maxwell theory arises in string theory
as the composite of
two pairs of electric and magnetic charges of various
fields of ten-dimensional string theory compactified to
four dimensions. To a distant observer, however, the composite
appears as a single black hole. 

Before proceeding with the entropy comparison for this
black hole, let us return to the elementary particle/black
hole correspondence to make sure we are on the right track.  
We first consider solutions which correspond to pure
states, or which have zero entropy. Such solutions 
necessarily have zero area. It turns out that this
is the case for those solutions which possess no
more than two constituents. For the two cases, 
the quantum numbers
(mass and charge) of the solutions
match those of particular supersymmetric quantum
string states. 
In addition, the dynamics of the black holes agree with the
four-point amplitudes of the corresponding string states 
in the low-velocity limit \cite{dynam}. For the single-charge
black holes and their corresponding states, this scattering
is trivial, while for the two-charge black holes and their 
corresponding states, we obtain Rutherford scattering. 
Of course this quantum number matching
and dynamical agreement does not
mean we can go ahead and identify the black hole
solutions with elementary string states, but 
at least the correspondence makes sense.

Let us now return to our black hole with $S \neq 0$,
or $A\neq 0$. This solution should
correspond to an {\em ensemble} of string states. Now
the laws of black hole thermodynamics follow from classical
general relativity. However, the laws of thermodynamics in 
general
follow from a microcanonical counting of quantum states,
{\em i.e.},  from statistical mechanics. An important test of 
a theory of quantum gravity is then the following: can one 
obtain the black hole laws of thermodynamics from a 
counting of microscopic quantum states, {\em i.e.}, is there
a quantum/statistical mechanical basis for these classical
laws? We are interested in performing this test for string
theory, where we have established a correspondence
between classical solutions and  elementary and solitonic
states. On the one side, we can construct a
black hole solution,
compute its area and deduce the entropy from $S=A/4G$.
On the other side, we can set up the ensemble of
states corresponding to this solution, compute its degeneracy
and take the logarithm to obtain the entropy.

As we have already mentioned, the
Reissner-Nordstr\"om black hole is the bound
state of four constituent
single-charge black holes. Let $g_{MN}$ and $B_{MN}$
be the spacetime metric and antisymmetric tensor.
The four charges corresponding
to the four constituent black holes are given by 
(normalized to represent 
quantum-mechanical number operators): $Q_e$, electric
with respect to $B_{\mu\nu}$, $Q_m$, magnetic with respect to 
$B_{\mu\nu}$,
$N_e$, electric with respect to $g_{5\mu}$ and 
$N_m$, magnetic with respect to $g_{4\mu}$, 
where $x_4$ and $x_5$
are two compactified directions. 
Let us also simplify the picture slightly by setting $N_m=1$.

The classical solution has a nonzero area given by 
$A=8\pi G \sqrt{Q_e Q_m N_e}$, hence a Bekenstein-Hawking
entropy $S_{BH}=2\pi \sqrt{Q_e Q_m N_e}$. The setup of the 
corresponding
string states is the following.
We are interested in the case of large charges, 
corresponding to black hole solutions. For large $N_e$,
the number operator $N_e$ represents
the momentum of massless open strings going between
$Q_e$ electric charges and $Q_m$ magnetic charges. 
The total number of
bosonic modes is then given by $4Q_e Q_m$, since there is a
degeneracy associated with the extra $(6789)$ part of the 
compactified space. By supersymmetry, the number of fermionic
modes is also $4Q_eQ_m$.
This system is then like a 
$1+1$-dimensional gas of massless left 
moving particles with $4Q_eQ_m$ bosonic and fermionic species of
particles carrying total energy $N_e/R$, where $R$ is the 
radius of the circle. The number of such modes is 
given by $d(N) \sim \exp{2\pi \sqrt{Q_e Q_m N_e}}$, so that 
$S=\ln d(N)=2\pi \sqrt{Q_e Q_m N_e}=S_{BH}$, in agreement
with the area law. This is very exciting, as it is the first
time we can derive the area law from a quantum-mechanical
theory 
(string theory).
This result was first found for five-dimensional extremal
black holes \cite{exfive}\ and subsequently for 
four-dimensional
extremal black holes \cite{exfour}. 
Analogous results for near-extremal
black holes were also obtained \cite{nonex}, 
which seems to indicate that
this sort of factorization is not a property of supersymmetry 
alone,
although it is only for supersymmetric solutions that one can
invoke non-renormalization theorems to protect the counting
of states in going from the perturbative state-counting picture
to the nonperturbative black hole picture.

The correspondence between black holes and string states can
be understood as follows \cite{gary}. 
String states at level $N$ have
mass $M_s\sim N/l_s^2$ and entropy $S_s\sim\sqrt{N}$, 
where $l_s$ is the string scale. This picture is valid
provided the string coupling $g<<1$, where $g$ is
related to Newton's constant $G$ in four dimensions via
$G \sim g^2 \l_s^2$. 
Now for the black hole solution of the low-energy field theory 
limit of string theory, the mass and entropy 
are given by $M_{BH} \sim R_S/G$ and $S_{BH} \sim R_S^2/G$,
where $R_S$ is the Schwarzschild radius. The mass and entropy of 
the string states become of the same order precisely when
$R_S \sim l_s$, {\it i.e.} when the string scale becomes of the
order of the Schwarzschild radius. This happens when 
$g \sim N^{-1/4}$. For $N$ very large, $g$ is still very small
at this point. It then makes sense to refer to 
$\tilde g=g N^{1/4}$ as our effective coupling. For
$\tilde g < < 1$, we have perturbative string states.
At $\tilde g \sim 1$, the black hole forms, the
low-energy solution still being valid since $g$ is still
small. As we continue to turn up the coupling, the black hole 
picture continues to hold until we reach the region 
$g > > 1$, in which case a strong/weak coupling duality
map presumably takes the black hole back into a perturbative
state in the weak coupling limit of another string theory. 

Of course we still do not understand the precise mechanism
by which an ensemble of states turns
into a black hole. This and other questions
remain to be answered, as well as the formulation of a 
duality-manifest string theory. In this regard,
connections with the better-understood Yang-Mills
\cite{SW}\ duality seem most promising.

\section*{Acknowledgments}
Research supported by NSERC of Canada and Fonds FCAR du 
Qu\' ebec.

\end{document}